\documentclass[11pt]{article}
\usepackage{amssymb}
\usepackage[dvips]{graphicx}
\usepackage{amsmath}
\begin{document}
\begin{center}

{\LARGE\bf Full Investigation on the Dynamics of Power-Law Kinetic
Quintessence}

 \vskip 0.15 in
 $^\dag$Wei Fang$^{1, 2, 3}$, Hong Tu$^{1, 2}$, Ying Li$^4$, Jiasheng Huang$^3$, Chenggang Shu$^2$
 \\ \

\small {$^1$\textit{Department~of~Physics,~Shanghai~Normal~University,~100~Guilin~Rd.,~Shanghai,~200234,~P.R.China}\\
        $^2$\textit{The~Shanghai~Key~Lab~for~Astrophysics,~100~Guilin~Rd.,~Shanghai,~200234,~P.R.China}\\
        $^3$\textit{Harvard-Smithsonian~Center~for~Astrophysics,~60~Garden~St.,~Cambridge,~MA~02138,~USA}}\\
        $^4$\textit{College~of~Information~Technology,~Shanghai~Ocean~University, ~Shanghai,~201306,~P.R.China}
 \footnotetext{$\dag$ \ \ wfang@shnu.edu.cn, wfang@cfa.harvard.edu}

 \vskip 0.5  in \centerline{\bf Abstract} \vskip 0.2 in

\begin{minipage} {5.8in} {\hspace*{10pt}\small

We give a full investigation on the dynamics of power-law kinetic
quintessence $L(X, \phi)=V(\phi)(-X+X^2)$ by considering the
potential related parameter $\Gamma$($=\frac{V V''}{V'^2}$) as a
function of another potential parameter $\lambda$($=\frac{V'}{\kappa
V^{3/2}}$), which correspondingly extends the analysis of the
dynamical system of our universe from two-dimension to
three-dimension. Beside the critical points found in previous
papers, we find a new de-Sitter-like dominant
attractor(\textit{cp$6$}) and give its stable condition using the
center manifold theorem. For the dark energy dominant
solution(\textit{cp$6$} and \textit{cp$7$}), it could be
distinguished from canonical quintessence and tachyon models since
the sound speed $c_s^2=0$ or $c_s^2\ll 1$. For the scaling solution
(\textit{cp$8$}), it is very interesting that the sound speed
$c_s^2=1/5$ while it behaves as ordinary matter. We therefore point
out that the power-law kinetic quintessence should have different
signatures on cold dark matter power spectrum and cosmic microwave
background both at early time when this scalar field is an early
dark energy with $\Omega_\phi$ being non-negligible at high redshift
and at late time when it drives the accelerating expansion. We even
do not know whether there are any degeneracies of the impacts
between these two epoches. They are expected to be investigated in
future.

}
\end{minipage}
\end{center}
\newpage
\section{Introduction} Power-law kinetic quintessence is a kind of k-essence model, described by the lagrangian $L(X, \phi)=V(\phi)(-X+X^2)$.
It is firstly proposed in one version of k-inflation models\cite{1}.
It is shown that this kind of models with a higher-order
non-canonical kinetic terms instead of the help of potential terms
can also drive an inflationary evolution starting from rather
generic initial conditions. It can roll slowly from a high-curvature
initial phase, down to a low-curvature phase and can exit inflation
to end up being radiation-dominated, in a naturally graceful
manner\cite{1}. It appeared as a candidate of dark energy
model\cite{2} to address the late time accelerating expansion (see
for example, \cite{add21,3, 4, 5, 6}). It is very interesting to
investigate this kinetic driven quintessence since it could behavior
like a cosmological constant while the sound speed $c_s^2$ could
dramatically be far less then 1 or even equal zero\cite{7}. The dark
energy, with its sound speed being very small compared to the speed
of light(namely $c_s^2\ll1$), is referred as cold dark energy
\cite{7add1, 7add2}. This feature of low value sound speed is
distinguishable from the standard $\Lambda$CDM model which has a
purely non-clustering dark energy component or the quintessence
model with $c_s^2=1$. The effect of $c_s^2\rightarrow0$ is to
suppress the integrated Sachs-Wolfe(ISW) effect at large angular
scales because the dark energy component can cluster and then reduce
the decay of the gravitational potential that causes the ISW
effect\cite{7, 8, 9, 10}. The low value of sound speed can also
enhance the matter power spectrum  that the dark energy clustering
induces at large scales, and the closer is $c_s^2$ to the speed of
light the smaller is the effect\cite{11}. In addition, combining
cluster abundances with CMB background power spectra can distinguish
a true sound speed of 0.1 from 1 at $99\%$ confidence\cite{12}.
Power-law kinetic quintessence model was also investigated in the
context of a brane world\cite{13}.
\par The dynamics of the power-law kinetic quintessence with inverse
square potential $V(\phi)\propto \phi^{-2}$ had been investigated in
detail using a phase-space analysis to its critical points\cite{18}.
However, there may exist new critical points and correspondingly
have the new cosmological implication if the potential is not
restricted to the inverse square potential according to the previous
results\cite{14, 15}. We therefore need to study the dynamical
evolution of power-law kinetic quintessence beyond the inverse
square potential to get a full investigation of this model. This
full investigation of the dynamics is really important since the
evolution of the dark energy is essential both in the late time and
in the early time of the universe. If the dark energy may have a
fraction of the critical density $\Omega_{de}(z_{lss})\simeq
10^{-2}$ at the CMB last scattering surface rather than
$\Omega_{de}(z_{lss})\simeq 10^{-9}$, and satisfy two requirements
of $w$ being significantly different from $-1$ and $c_s^2\ll1$ at
that time, the perturbations in the dark energy will have an
appreciable influence on the matter power spectrum and large scale
clustering\cite{12}. It is well known that the scaling solution
satisfies these requirements with $0<\Omega_{de}<1$ and
$w_{de}=w_b$. In order to investigate the dynamics of the power-law
kinetic quintessence beyond the inverse square potential, we will
rely on the method described in\cite{14, 15, 16}. It helps us to
explore the critical points and the general dynamical behavior of
the power-law kinetic quintessence with nearly arbitrary potentials
rather than just one special potential. We will rely on the
three-dimensional dynamical autonomous systems for power-law kinetic
quintessence obtained in paper\cite{17}, try to find all the
critical points under the observable related variables $(w_{\phi},
\Omega_{\phi}, \lambda)$ instead of previous trivial variables $(x,
y, \lambda)$, and give the cosmological implication. The paper is
organized as follows. We firstly give the basic framework and the
three dimensional dynamical system in section 2, and then explore
the classical and quantum stabilities and present its constraint on
the value of $\gamma_\phi$ in section 3. We give all the critical
points of three dimensional dynamical system Eqs.(\ref{eq5},
\ref{eq6}, \ref{eq10}), analyze their existence and stable
conditions and investigate their cosmological properties in section
4. We finally find the differences of power-law kinetic quintessence
with cosmological constant, canonical quintessence and tachyon,
discuss the cosmological implications and give our conclusions in
section 5.
\section{Basic Framework and three Dimensional Dynamical System}

Let us restrict ourselves to a flat universe described by the FRW
metric, and consider a spatially homogeneous real scalar field
$\phi$ with non-canonical kinetic energy term. The lagrangian
density is given as

\begin{equation}\label{eq1}p_{\phi}=L(X, \phi)=V(\phi)(-X+X^2)\end{equation}
where
$X=\frac{1}{2}\nabla_{\mu}\phi\nabla^{\mu}\phi=\frac{1}{2}{\dot{\phi}}^2$
for a spatially homogeneous scalar field.
 The pressure, energy density and sound speed of the scalar field could be easily obtained
as the following:

\begin{equation}\label{eq2}\rho_{\phi}=2X\frac{\partial p}{\partial X}-p=V(\phi)(-X+3X^2)\end{equation}
\begin{equation}\label{eq3}H^2=(\frac{\dot{a}}{a})^2=\frac{1}{3M^2_{pl}}[\rho_{\phi}+\rho_b]\end{equation}
\begin{equation}\label{eq4}\dot H=-\frac{1}{2M^2_{pl}}[2V(\phi)(-X+2X^2)+\gamma_b\rho_b]\end{equation}
where $8\pi G=\kappa^2=1/M^2_{pl}$, $\rho_b$ is the density of a
barotropic fluid component with the equation of state $p_b=w_b
\rho_b=(\gamma_b-1) \rho_b$. $\gamma_b=1$ for matter and
$\gamma_b=4/3$ for radiation.

\par The three dimensional autonomous dynamical system is given as follows\cite{17}:

\begin{equation}\label{eq5}\frac{d \Omega_{\phi}}{dN}=f_1(\Omega_{\phi}, \gamma_{\phi}, \lambda)=3(\gamma_b-\gamma_{\phi})\Omega_{\phi}(1-\Omega_{\phi})\end{equation}
\begin{equation}\label{eq6}\frac{d \gamma_{\phi}}{dN}=f_2(\Omega_{\phi}, \gamma_{\phi}, \lambda)=\frac{(\lambda\sqrt{3(4-3\gamma_{\phi})\Omega_{\phi}}+3\gamma_{\phi})(\gamma_{\phi}-2)(3\gamma_{\phi}-4)}{3\gamma_{\phi}-8}\end{equation}
\begin{equation}\label{eq7}\frac{d\lambda}{dN}=\lambda^2\sqrt{3(4-3\gamma_{\phi})\Omega_{\phi}}~(\Gamma-\frac{3}{2})\end{equation}
 where
\begin{equation}\label{eq8}\lambda=\frac{V'}{\kappa V^{3/2}}, ~\Gamma=\frac{V V''}{V'^2},~\Omega_{\phi}=\frac{\rho_{\phi}}{3M^2_{pl}H^2},\end{equation}
The equation of state $w_{\phi}$ and the sound speed $c_s^2$ of dark
energy are as follows:
\begin{equation}\label{eq81}w_{\phi}=\gamma_{\phi}-1=\frac{X-1}{3X-1},~c_s^2=\frac{p_{,X}}{\rho_{,X}}=\frac{2X-1}{6X-1}=\frac{-\gamma_\phi}{3\gamma_\phi-8}\end{equation}

Above Eqs.(\ref{eq5}-\ref{eq7}) completely describe the dynamical
evolution of the power-law kinetic quintessence. Eq.(\ref{eq7}) will
vanish when $\Gamma=3/2$, then the dynamical system
Eqs.(\ref{eq5}-\ref{eq7}) will reduce to a two dimensional
autonomous system which corresponds to the inverse square potential
$V(\phi)=(\frac{1}{2}\kappa \lambda \phi-c_1)^{-2}$. Authors had
obtained this two-dimensional dynamical autonomous system with the
dimensionless variables $(x, y)$, and studied the phase-space
properties and the cosmological implications of the critical points
in detail. However, here we give the two-dimensional autonomous
system Eqs.(\ref{eq5}-\ref{eq6}) with the variables being
observation related quantities
 $(\Omega_{\phi}, \gamma_{\phi})$ instead of $(x, y)$. We will obtain the critical points of the
observational quantities $(\Omega_{\phi}, \gamma_{\phi})$ directly,
so it will be more convenient to study the properties of the
critical points and their cosmological implications with these new
variables. Furthermore, we will investigate the dynamics of the
three dimensional dynamical system instead of the two dimensional
dynamical system, and correspondingly, we can study the dynamics of
power-law kinetic quintessence beyond the inverse square potential.
We will get to know which critical points are the critical points
for all the power-law kinetic quintessence(no matter with the form
of the potentials) and which are only relative to the concrete
potentials. We rely on the method which is proposed in Refs\cite{14,
16} and then generalized to several other cosmological
contexts\cite{15}\cite{20}-\cite{31}.
\par We briefly introduce the idea of our treatment here. When the potential is not the inverse
square potential, the potential related parameter
$\Gamma\neq\frac{3}{2}$. In this case, another potential related
parameter $\lambda$ is a dynamically changing quantity, then the
system Eqs.(\ref{eq5}-\ref{eq7}) will be not an autonomous system
any more since $\Gamma$ is unknown, and therefore we can not analyze
the phase space like the inverse square potential exactly. However,
since $\lambda$ is the function of tachyon field $\phi$ and $\Gamma$
is also the function of $\phi$, so $\Gamma$ can be expressed as a
function of $\lambda$ in principle:

\begin{equation}\label{eq9}\Gamma(\lambda)=f(\lambda)+\frac{3}{2}\end{equation}
then Eq.(\ref{eq7}) becomes:
\begin{equation}\label{eq10}\frac{d\lambda}{dN}==f_3(\Omega_{\phi}, \gamma_{\phi}, \lambda)=\lambda^2\sqrt{3(4-3\gamma_{\phi})\Omega_{\phi}}~f(\lambda)\end{equation}

Obviously, Eqs.(\ref{eq5}-\ref{eq6}) and Eq.(\ref{eq10}) are a
dynamical autonomous system. Given any form of the function
$f(\lambda)$, we can get the corresponding exact expression for the
potential $V(\phi)$(see Ref\cite{15} for details).  The
three-dimension autonomous system Eqs.(\ref{eq5}, \ref{eq6},
\ref{eq10}) reduces to two-dimension autonomous systems when
$f(\lambda)=0$(i.e, $V(\phi) \propto \phi^{-2}$, $\Gamma=3/2$ and
$\lambda=constant$).
\par Let us focus on Eq.(\ref{eq10}) to show you why we state that studying the dynamics based on three dimensional
system is far superior to two dimensional system. \textit{Firstly},
all the critical points obtained in two dimensional system when the
potential $V(\phi) \propto \phi^{-2}$ is just the special case when
$f(\lambda_\ast)=0$, where $\lambda_\ast$ is the value that makes
$f(\lambda_\ast)=0$. We should keep in mind that there are many
potentials with their $f(\lambda)$ could be zero, the inverse square
potential $V(\phi) \propto \phi^{-2}$ is just the simplest case. For
example, $V(\phi)=V_0/(\phi^2-\phi_0^2)$ corresponds to
$f(\lambda)=1/2-2/(V_0\kappa^2\lambda^2)$\cite{15}.
$\lambda_\ast=\pm 2/(\kappa\sqrt{V_0})$ makes $f(\lambda_\ast)$
equal $0$. That means all the critical points exist for inverse
square potential will also exist for the potential
$V(\phi)=V_0/(\phi^2-\phi_0^2)$. Obviously, these critical points
will not exist for the exponential potential $V(\phi)=V_0 e^{\alpha
\phi}$ since in this case $f(\lambda)$ does not equal $0$ (it always
equals $-1/2$). \textit{Secondly}, we can find the new critical
points which will not exist for inverse square potential. We can
easily understand it from Eq.(\ref{eq10}). Generally speaking, there
are four possibilities to make Eq.(\ref{eq10}) $d\lambda/dN=0$:
$f(\lambda)=0$, $\lambda=0$, $\gamma_{\phi}=4/3$ and
$\Omega_{\phi}=0$. We have discussed earlier in this paragraph about
the case of $f(\lambda)=0$. We should emphasize that the last three
types of critical points exist even if the potential is not the
inverse square potential. For the second case $\lambda=0$, the
potentials with an extremum(i.e., $V'=0$) possess these critical
points of $\lambda=0$ since the potential related parameter
$\lambda=V'/(\kappa V^{3/2})$. In fact, not only the potentials with
an extremum but all those potentials with $\lambda$ being zero in
function $f(\lambda)$ have these critical points. For example,
$f(\lambda)=\beta\lambda-\frac{1}{2}$, the corresponding potential
has an implicit expression as $2\beta
V(\phi)^{-\frac{1}{2}}-\frac{1}{2}c_1 ln(V(\phi))=-\frac{1}{2}\kappa
\phi+c_2$\cite{15}. However, for the potential
$V(\phi)=V_0/(\phi^2-\phi_0^2)$ as previously mentioned in this
paragraph, there is no such critical points since
$f(\lambda)=1/2-2/(V_0\kappa^2\lambda^2)$, therefore $\lambda$ can
not be zero. For the last two cases $\gamma_{\phi}=4/3$ and
$\Omega_{\phi}=0$, the corresponding critical points even exist
irrespective of the potentials. We will give all the critical points
and analyze their properties in detail in Section 4.

\section{Classical and Quantum Stabilities}
Before we investigate the critical points and their properties, we
consider the range of $\gamma_{\phi}$ as well as the classical and
quantum stabilities of the power-law kinetic quintessence.
\par
\begin{figure}[htb]
\begin{minipage}[t]{0.48\linewidth}
\centering
\includegraphics[scale=0.41,origin=c,angle=270,bb=118 95 533 660,clip]{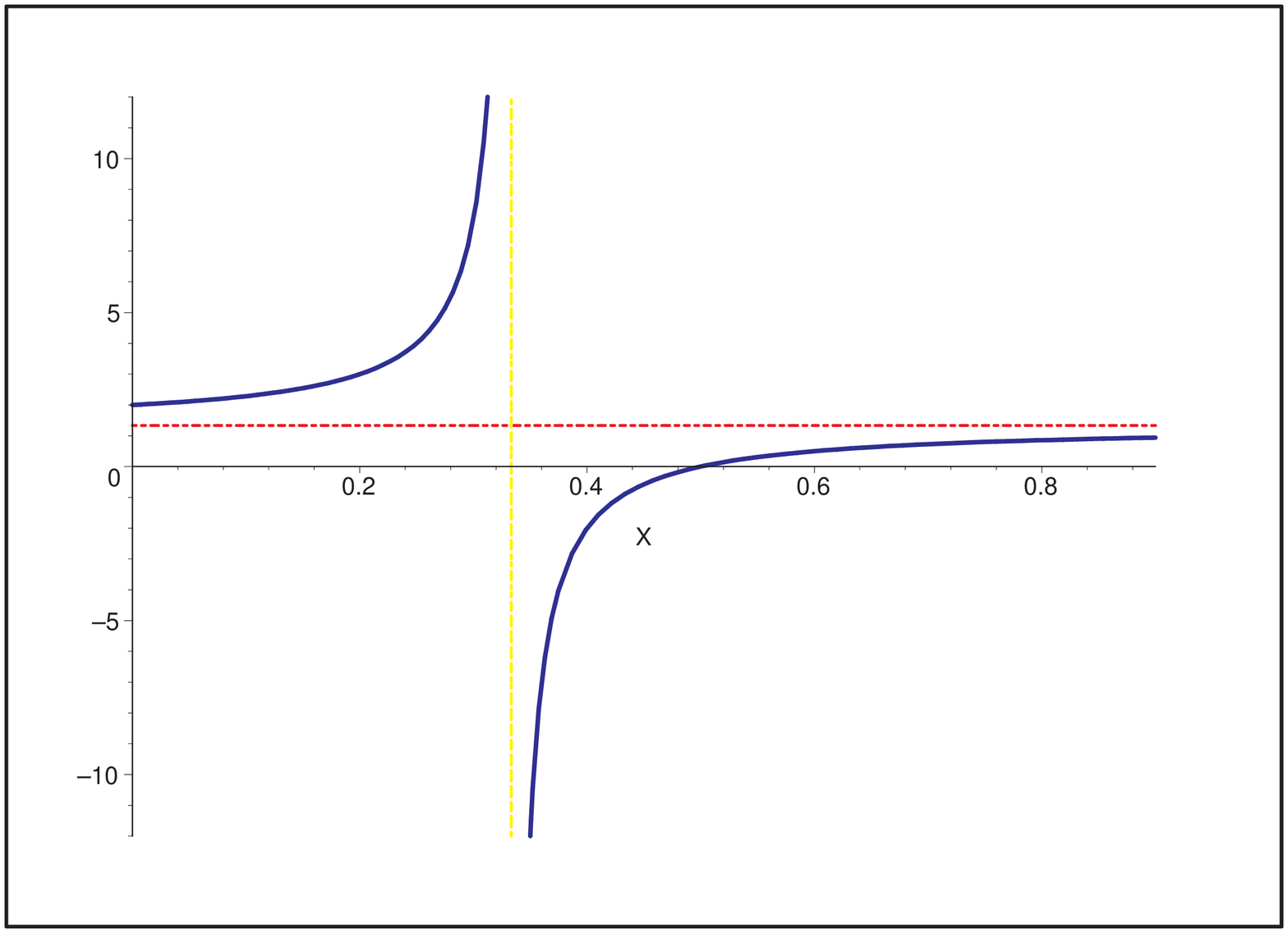}
 \caption{The evolution of $\gamma_{\phi}$ with respect to $X$. The yellow dash vertical line is $X=1/3$, where $\gamma_{\phi}\rightarrow\pm\infty$.
 The red dash horizontal line is $\gamma_\phi=4/3$, $\gamma_\phi\rightarrow4/3$ when $X\rightarrow\infty$.
 $0\leq\gamma_{\phi}\leq4/3$ is required from the classical and quantum stabilities. } \label{fig1}
\end{minipage}
\hfill
\begin{minipage}[t]{0.48\linewidth}
\centering
\includegraphics[scale=0.41,origin=c,angle=270,bb=118 95 533 660,clip]{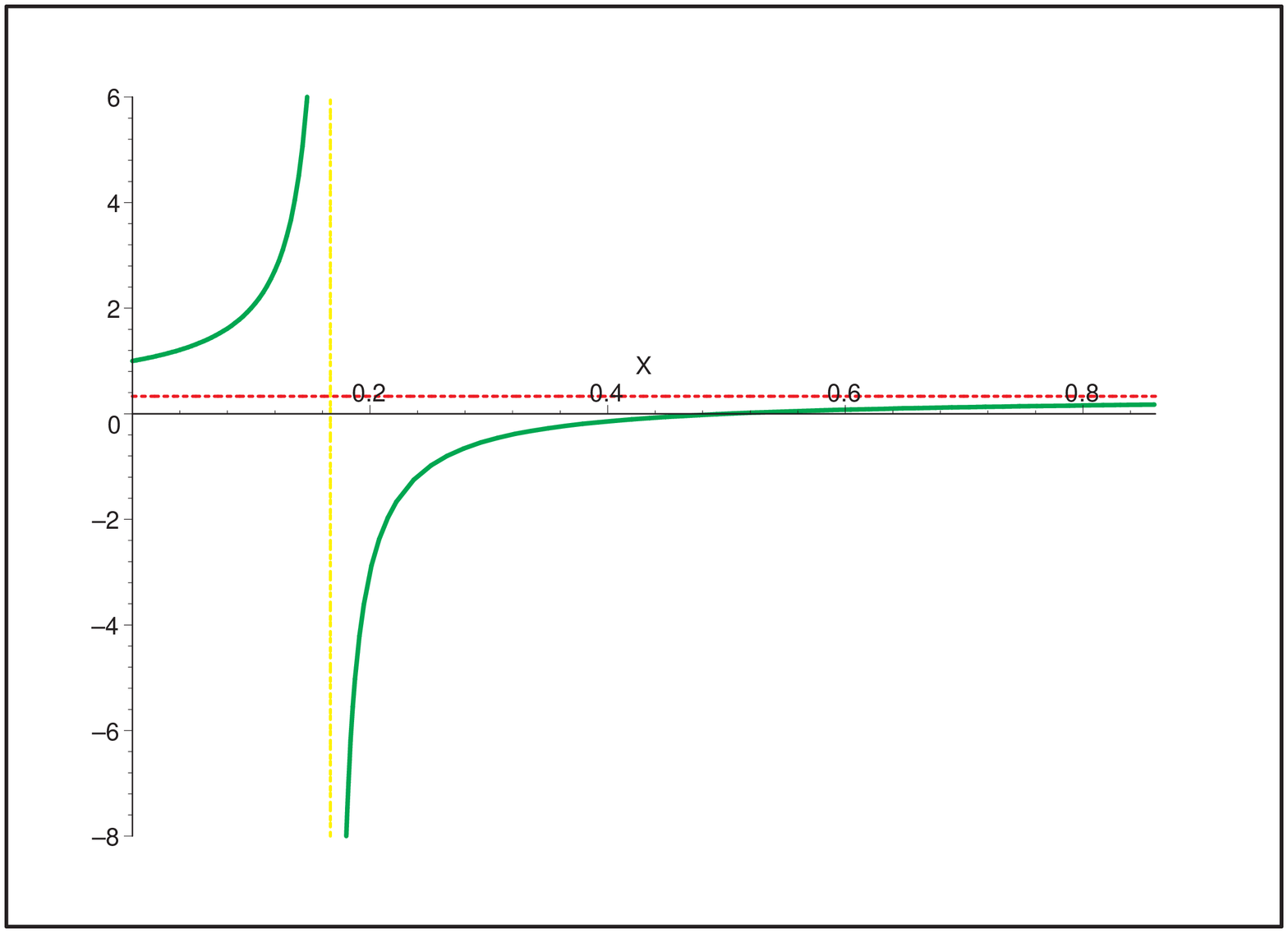}
\caption{The evolution of $c_s^2$ with respect to $X$. The yellow
dash vertical line is $X=1/6$, where $c_s^2\rightarrow\pm\infty$.
 The red dash horizontal line is $c_s^2=1/3$, $c_s^2\rightarrow1/3$ when $X\rightarrow\infty$.} \label{fig2}
\end{minipage}
\end{figure}

 We get the expression $\gamma_{\phi}=1+w_{\phi}=(4X-2)/(3X-1)$ from Eq.(\ref{eq81}). Since $X=\dot{\phi}^2/2>0$, we can easily
get the range of  $\gamma_{\phi}$:  $\gamma_{\phi}\geq 2$ or
$\gamma_{\phi}<4/3$. We plot the evolution of $\gamma_{\phi}$ with
respect to $X$ in Fig.\ref{fig1}. However, there are two constraints
on the value of $X$ if we consider the classical and quantum
stabilities. When we consider the stability of classical
perturbations, the sound speed $c_s^2$ should be positive. We
therefore get that $0\leq X<1/6$ or $X\geq1/2$. We plot the
evolution of $c_s^2$ with respect to $X$ in Fig.\ref{fig2}. If we
consider the quantum stability, which requires that the perturbed
Hamiltonian about a background solution is positive, demanding that
$P_{,X}\geq0, P_{,X}+2XP{,XX}\geq0$. That leads to $2X-1\geq0$ and
then we have $X \geq 1/2$. So finally we obtain that $X \geq 1/2$
for the stability from both classical and quantum points of view.
What we discuss here can also be found in detail in Refs\cite{18,
32, 33}. We therefore get the range for $\gamma_{\phi}$:

\begin{equation}\label{eq11}0\leq\gamma_{\phi}<4/3 ~~~ (-1 \leq w_{\phi}< 1/3)\end{equation}

Obviously, we get above equation Eq.(\ref{eq11})(see Fig 1) from the
requirements of the classical and  quantum stabilities. However, the
interesting thing is that, we can easily get the similar constraint
from three dimensional dynamical system Eq.(\ref{eq6}) or
Eq.(\ref{eq7}) since there is a term $\sqrt{3(4-3\gamma_{\phi})}$.
We do not know it is just occasional or there are some reasons that
make the dynamical system give the similar constraint.

\section{Critical Points and their Cosmological Properties}
\par We will investigate the critical points and their properties in this section.
The critical points can be found by setting
$d\Omega_{\phi}/dN=d\gamma_{\phi}/dN=d\lambda/dN=0$ while their
properties are determined by the eigenvalues of the Jacobi matrix
${\cal A}$ of the three dimensional nonlinear autonomous system
Eqs.(\ref{eq5}, \ref{eq6}, \ref{eq10}). The Jacobi matrix ${\cal A}$
 of each point is obtained by linearizing the three
dimensional nonlinear autonomous system Eqs.(\ref{eq5}, \ref{eq6},
\ref{eq10}) around each critical point\cite{14},
\begin{equation}\label{eqA}{\cal A}^\S=\left[ \begin{array}{lll}
\partial f_1(\Omega_{\phi}, \gamma_{\phi}, \lambda)/\partial \Omega_{\phi} & \partial f_1(\Omega_{\phi}, \gamma_{\phi}, \lambda)/\partial \gamma_{\phi} &\partial f_1(\Omega_{\phi}, \gamma_{\phi}, \lambda)/\partial \lambda \\
\partial f_2(\Omega_{\phi}, \gamma_{\phi}, \lambda)/\partial \Omega_{\phi} & \partial f_2(\Omega_{\phi}, \gamma_{\phi}, \lambda)/\partial \gamma_{\phi} &\partial f_2(\Omega_{\phi}, \gamma_{\phi}, \lambda)/\partial \lambda \\
\partial f_3(\Omega_{\phi}, \gamma_{\phi}, \lambda)/\partial \Omega_{\phi} & \partial f_3(\Omega_{\phi}, \gamma_{\phi}, \lambda)/\partial \gamma_{\phi} &\partial f_3(\Omega_{\phi}, \gamma_{\phi}, \lambda)/\partial \lambda
\end{array}\right]_{each~critical~point}\end{equation}

\footnotetext{$\S$ \textit{Since there are terms of
$\sqrt{\Omega_{\phi}}$ and $\sqrt{4-3\gamma_{\phi}}$ in dynamical
system Eqs.(\ref{eq5}, \ref{eq6}, \ref{eq10}), some elements in
Jacobi matrix ${\cal A}$ will diverge to $\infty$ for cp1, cp3 and
cp4. For these kind of critical points, we used the method of
substitution to remove the square roots and got the well-defined
Jacobi matrix ${\cal A}$}.}

\par \textit{Critical point 1}(hereafter \textit{cp1}): $(\Omega_{\phi}=0, \gamma_{\phi}=0,
\lambda=\lambda_{ar})$($\lambda_{ar}$ means an arbitrary real
constant). \textit{Cp1} always exists independent of the form of the
potential. It is a barotropic fluid dominated solution. However,
since the eigenvalues of \textit{cp1} found from its Jacobi matrix
is $(3\gamma_b/2, -3, 0)$, so it is an unstable saddle point.

\par \textit{Critical point 2} (hereafter \textit{cp2}): $(\Omega_{\phi}=0, \gamma_{\phi}=2,
\lambda=\lambda_{ar})$. This point actually does not exist since
$\gamma_\phi=2>4/3$.

\par \textit{Critical point 3}(hereafter \textit{cp3}): $(\Omega_{\phi}=0, \gamma_{\phi}=4/3,
\lambda=\lambda_{ar})$. This point always exists independent of the
form of the potential. It is also a barotropic fluid dominated
solution. The eigenvalues of \textit{cp3} is $(3\gamma_b/2-2, 1,
0)$, so it is an unstable saddle point(for barotropic fluid being
matter, $\gamma_b=1$) or an unstable node point(for barotropic fluid
being radiation, $\gamma_b=4/3$).

\par \textit{Critical point 4}(hereafter \textit{cp4}): $(\Omega_{\phi}=1, \gamma_{\phi}=4/3,
\lambda=\lambda_{ar})$. This point always exists independent of the
form of the potential. \textit{cp4} is a power-law kinetic
quintessence dominated solution$(\Omega_{\phi}=1)$ where this scalar
field behaves as radiation$(\gamma_{\phi}=4/3)$. However, it is an
unstable node point(whatever barotropic fluid is matter or
radiation) since the eigenvalues of \textit{cp4} is $(4-3\gamma_b,
1, 0)$.

\par \textit{Critical point 5}(hereafter \textit{cp5}): $(\Omega_{\phi}=1, \gamma_{\phi}=2,
\lambda_1)$, where $\lambda_1=0$ or $f(\lambda_1)=0$. However, this
point does not exist since the value of $\gamma_\phi$ should be
$0\leq\gamma_\phi<4/3$.

\par \textit{Critical point 6}(hereafter \textit{cp6}): $(\Omega_{\phi}=1, \gamma_{\phi}=0,
\lambda=0)$. This point exists depending on the form of the
potential. All the potentials with $\lambda$ being zero in function
$f(\lambda)$ have \textit{cp6}. For example,
$f(\lambda)=\beta\lambda-\frac{1}{2}$, the corresponding potential
has an implicit expression as $2\beta
V(\phi)^{-\frac{1}{2}}-\frac{1}{2}c_1 ln(V(\phi))=-\frac{1}{2}\kappa
\phi+c_2$\cite{15}. However, for the potential
$V(\phi)=V_0/(\phi^2-\phi_0^2)$, there is no such critical point
since $f(\lambda)=1/2-2/(V_0\kappa^2\lambda^2)$, therefore $\lambda$
can not be zero. For the potentials with an extremum(i.e., $V'=0$,
for example, $V(\phi)=m\phi^2/2+V_0$) also possess this critical
point since the potential related parameter $\lambda=V'/(\kappa
V^{3/2})$. \textit{Cp6} is very interesting since it corresponds to
the universe dominated by the dark energy which behaves as an
cosmological constant with the sound speed $c_s^2$ being 0.
Furthermore, \textit{cp6} could be a stable point since the
eigenvalues of this point is $(-3\gamma_b, -3, 0)$. If a critical
point of a linear three dimensional dynamical system has the
eigenvalues $(-3\gamma_b, -3, 0)$, we can state directly that it is
a stable point even though one of the eigenvalues equals zero.
However, the dynamical system Eqs.(\ref{eq5},\ref{eq6}, \ref{eq10})
in our paper is a nonlinear system, the eigenvalues $(-3\gamma_b,
-3, 0)$ is not enough to determine its stability\cite{34}. We need
to pursue the stable condition using center manifold
theorem\cite{14,34}. Appendix gives the detailed process to find the
stable condition using the center manifold theorem. We find that the
stable condition for \textit{cp6} is $f(\lambda)|_{\lambda=0}<0$,
i.e., $f(0)<0$.

\par \textit{Critical point $7$}(hereafter \textit{cp$7$}):
$(\Omega_{\phi}=1,
\gamma_{\phi}=-(3\lambda_\ast^3\pm\lambda_\ast\sqrt{9\lambda_\ast^2+48}),
\lambda=\lambda_\ast)$

\par \textit{Critical point $7_1$}(hereafter \textit{cp$7_1$}):
$(\Omega_{\phi}=1,
\gamma_{\phi}=-(3\lambda_\ast^3+\lambda_\ast\sqrt{9\lambda_\ast^2+48}),
\lambda=\lambda_\ast)$, where $\lambda_\ast$ makes
$f(\lambda_\ast)=0$. When $-0.508 \leq\lambda_\ast\leq 0$, we have
$0\leq\gamma_\phi<2/3$(see Fig.\ref{fig3}), so it is the condition
for accelerating expansion. This critical point exist for all the
potentials with their $f(\lambda_\ast)=0$. The simplest case is the
inverse square potential $V(\phi) \propto \phi^{-2}$, which is
considered in two dimensional system. However, there are many
potentials with their $f(\lambda)$ could be zero, the inverse square
potential $V(\phi) \propto \phi^{-2}$ is just the simplest case. We
can take $V(\phi)=V_0/(\phi^2-\phi_0^2)$ corresponding to
$f(\lambda)=1/2-2/(V_0\kappa^2\lambda^2)$\cite{15} as an example.
$\lambda_\ast=\pm 2/(\kappa\sqrt{V_0})$ when $f(\lambda_\ast)=0$.
That means all the critical points exist for inverse square
potential will also exist for the potential
$V(\phi)=V_0/(\phi^2-\phi_0^2)$. However, these critical points will
not exist for the exponential potential $V(\phi)=V_0 e^{\alpha
\phi}$ since in this case $f(\lambda)$ always equals $-1/2$. The
value of $\gamma_\phi$ and the eigenvalues of this critical point is
quite complicated, we will give its existence and stable condition
using numerical analysis. Considering the constraint Eq.(\ref{eq11})
from the requirements of the classical and quantum stabilities, the
existence condition is $-0.842 <\lambda_\ast\leq 0$. The stable
condition is $-0.662 \leq\lambda_\ast\leq 0$ and
$df_{\lambda_\ast}<0$ for both $\gamma_b=1$ and $\gamma_b=4/3$,
where $df_{\lambda_\ast}$ is the value of
$df(\lambda)/d\lambda|_{\lambda_\ast}$. We found that the condition
for accelerating expansion $-0.508 \leq\lambda_\ast\leq 0$ lies in
the range of stable condition, therefore the accelerating expansion
could be a stable solution.

\par \textit{Critical point $7_2$}(hereafter \textit{cp$7_2$}):
$(\Omega_{\phi}=1,
\gamma_{\phi}=-(3\lambda_\ast^3-\lambda_\ast\sqrt{9\lambda_\ast^2+48}),
\lambda=\lambda_\ast)$. When $-1.923\leq\lambda_\ast\leq -1.692$ or
 $0\leq\lambda_\ast\leq 0.689$ or
 $1.247\leq\lambda_\ast\leq 1.692$, we have
$0\leq\gamma_\phi<2/3$(see Fig.\ref{fig3}), so it is the condition
for accelerating expansion. The existence condition is
$-2.095\leq\lambda_\ast\leq -1.692$ or $0\leq\lambda_\ast\leq
1.692$. If $\gamma_b=1$, the stable condition is $-2.014
\leq\lambda_\ast\leq -1.692 ~and~ df_{\lambda_\ast}<0$ or
$0\leq\lambda_\ast\leq 0.328 ~and~ df_{\lambda_\ast}<0$.  If
$\gamma_b=4/3$, the stable condition is $-2.065 \leq\lambda_\ast\leq
-1.692 ~and~ df_{\lambda_\ast}<0$ or $0\leq\lambda_\ast\leq 0.328
~and~ df_{\lambda_\ast}<0$. We found the range of the accelerating
condition is overlapped with the stable condition, so
$\textit{cp$7_2$}$ could be a stable accelerating expansion
solution. The condition for a stable solution with an accelerating
expansion is as follows: $-1.923 \leq\lambda_\ast\leq -1.692 ~and~
df_{\lambda_\ast}<0$ or $0\leq\lambda_\ast\leq 0.328 ~and~
df_{\lambda_\ast}<0$.

\par The property of \textit{cp$7_2$} is very similar to
\textit{cp$7_1$}, we therefore refer to \textit{cp$7$} as
\textit{cp$7_1$} and \textit{cp$7_2$}. We are very interested in
these two points since they both could be a stable solution with
$\Omega_{\phi}=1$ and $w_{\phi}$ being any value between $-1$ and
$-1/3$.  Both \textit{cp6} and \textit{cp$7$} could be a stable
solution with dark energy dominating our universe. However
\textit{cp$7$} is more interesting than \textit{cp6} since the state
equation $w_{\phi}$ of \textit{cp$7$} could be any value between
$-1$ and $-1/3$. We find that there is no overlap among the
existence condition of \textit{cp6}, \textit{cp$7_1$} and
\textit{cp$7_2$}, so they will not exist simultaneously.

\par \textit{Critical point $8$}(hereafter \textit{cp$8$}):
$(\Omega_{\phi}=3\gamma_b^2/[(4-3\gamma_b)\lambda_\ast^2],
\gamma_{\phi}=\gamma_b, \lambda=\lambda_\ast)$, where $\lambda_\ast$
makes $f(\lambda_\ast)=0$. This critical point exist for all the
potentials with their $f(\lambda_\ast)=0$, similar to the critical
points \textit{cp$7_1$} and \textit{cp$7_2$}. If barotropic fluid is
radiation(i.e., $\gamma_b$=4/3), \textit{cp$8$} have no meaning
since
$\Omega_{\phi}=3\gamma_b^2/[(4-3\gamma_b)\lambda_\ast^2]\rightarrow\infty$$^\ddagger$.
\footnotetext{$\ddagger$ \textit{We can study this special case
directly from the dynamical system Eqs.(\ref{eq5}, \ref{eq6},
\ref{eq10}). When $\gamma_\phi=\gamma_b=4/3$,
$d\Omega_\phi/dN=d\gamma_\phi/dN=d\lambda/dN=0$ for any value of
$\Omega_{\phi}$. This critical point exists independent of the form
of the potentials. We can find the eigenvalues for this point is
$(0, 0, 1)$, so it is an unstable point. In fact, we find that this
point is actually a special case of \textit{cp4} with
$\gamma_b=4/3$.}}
 So here barotropic fluid could only be matter(i.e.,
$\gamma_b$=1), then $\Omega_{\phi}=3/\lambda_\ast^2$, we therefore
obtain $|\lambda_\ast|\geq\sqrt3$ to make $\Omega_{\phi}\leq 1$. The
stable condition for this scaling solution is $\lambda_\ast\leq
-\sqrt{3}$ and $df_{\lambda_\ast}<0$(stable node for
$-2\sqrt{2}\leq\lambda_\ast\leq -\sqrt{3}$ and stable spiral for
$\lambda_\ast< -2\sqrt{2}$). \textit{Cp$8$} is the scaling solution
where neither the scalar field nor the ordinary matter entirely
dominates the universe. The scalar field behaves as the ordinary
matter in this case.

\section{Discussions and Conclusions}
We have found all the critical points of three dimensional dynamical
nonlinear autonomous system of power-law kinetic quintessence, given
their existence and stable conditions and analyzed their
cosmological implications in section 4. Here we will give the
discussions and conclusions about their cosmological implications.

\par There are totally eight critical points for the dynamical system Eqs.(\ref{eq5}, \ref{eq6}, \ref{eq10}),
but only six critical points(\textit{cp$1$}, \textit{cp$3$},
\textit{cp$4$}, \textit{cp$6$} ,\textit{cp$7$}, \textit{cp$8$})
exist if the classical and quantum stability was considered. Among
these critical points, \textit{cp$1$}, \textit{cp$3$} and
\textit{cp$4$} always exist and are independent of the form of the
potentials. However, all of these critical points which are
independent of potentials are unstable. \textit{cp$1$} is an
unstable saddle point while \textit{cp$4$} is an unstable node.
\textit{cp$3$} is quite special since it is an unstable saddle point
when $\gamma_b=1$(matter) and an unstable node point when
$\gamma_b=4/3$(radiation). From a mathematical point of view, the
unstable node is different from the unstable saddle. All the
trajectories of an unstable node will move away from the critical
point to infinite-distant away while some trajectories of a saddle
are drawn to the critical point and other trajectories recede. So
the unstable node and saddle actually have different cosmological
implication even though they are all unstable.
\begin{figure}[htb]
\begin{minipage}[t]{0.48\linewidth}
\centering
\includegraphics[scale=0.38,origin=c,angle=270,bb=118 95 533 660,clip]{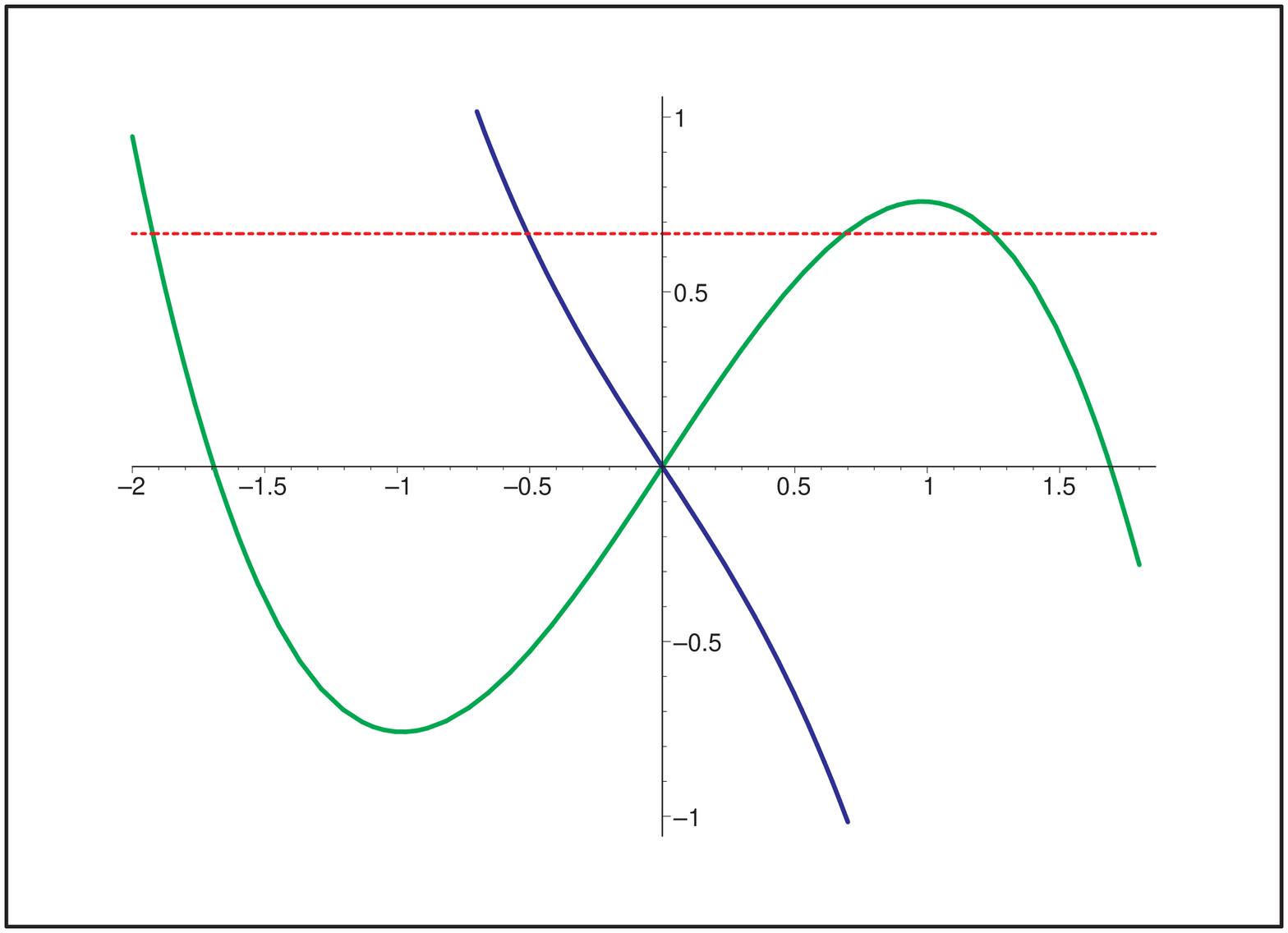}
 \caption{The evolution of $\gamma_{\phi}$ with respect to $\lambda_\ast$. Blue line is for \textit{cp$7_1$}
 with $\gamma_{\phi}=-(3\lambda_\ast^3+\lambda_\ast\sqrt{9\lambda_\ast^2+48})$ while solid line is for \textit{cp$7_2$}
 with $\gamma_{\phi}=-(3\lambda_\ast^3-\lambda_\ast\sqrt{9\lambda_\ast^2+48})$. Red dash horizontal line is $\gamma_\phi=2/3$.
 $0\leq\gamma_{\phi}\leq 2/3$ is required for the accelerating expansion of universe.} \label{fig3}
\end{minipage}
\hfill
\begin{minipage}[t]{0.48\linewidth}
\centering
\includegraphics[scale=0.38,origin=c,angle=270,bb=118 95 533 660,clip]{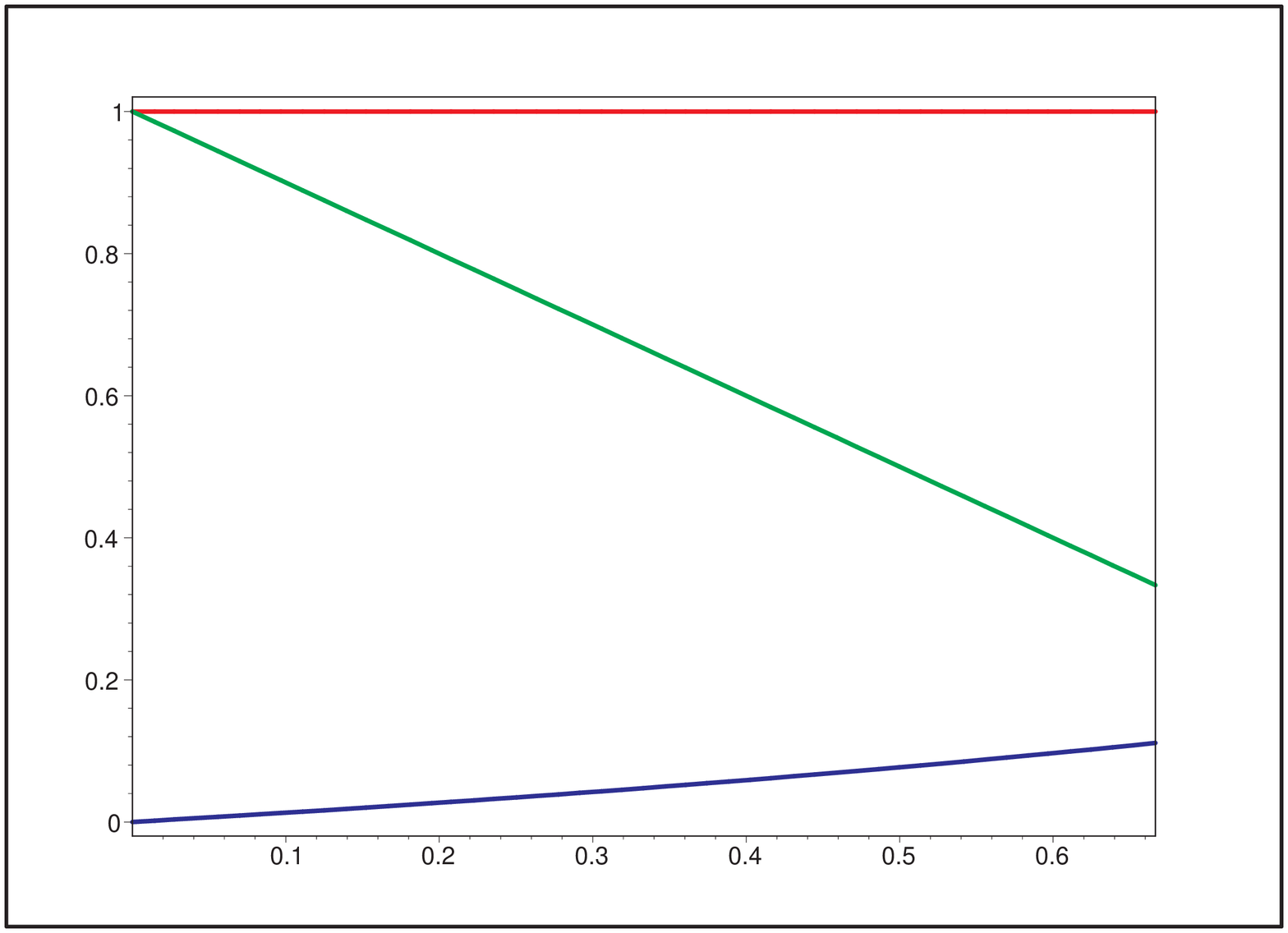}
\caption{The evolution of the sound speed $c_s^2$ with respect to
$\gamma_\phi$. Blue line is for power-law kinetic
quintessence($c_s^2=-\gamma_\phi/(3\gamma_\phi-8)$), where $c_s^2$
is monotonically increasing from $0$ to $1/9$ when $\gamma_\phi$
varies from $0$ to $2/3$. Red line is for quintessence($c_s^2=1$),
green line is for tachyon($c_s^2=1-\gamma_\phi$).} \label{fig4}
\end{minipage}
\end{figure}

\par The existence of other three critical points \textit{cp$6$},
\textit{cp$7$} and \textit{cp$8$} depend on the form of the
potentials. It is very interesting that all these three critical
points could be stable points if some conditions are satisfied.
\textit{Cp$6$} exists when the potential related parameter $\lambda$
could be $0$ in the function $f(\lambda)$. \textit{Cp$7$} and
\textit{cp$8$} exists when $f(\lambda)$ could be $0$. These three
critical points \textit{cp$6$}, \textit{cp$7$} and \textit{cp$8$}
have more important cosmological implication than \textit{cp$1$},
\textit{cp$3$} and \textit{cp$4$}. \textit{Cp$6$} is a new critical
point which is found only in three dimensional dynamical system.
\textit{Cp$6$} corresponds to the dark energy dominated
universe($\Omega_\phi=1$) where power-law kinetic quintessence
behaves as an cosmological constant with the sound speed $c_s^2$
being 0. So \textit{cp$6$} is a little different from canonical
quintessence and tachyon model since $c_s^2=1$ for both of those
scalar field. \textit{Cp$7$} and \textit{cp$8$} correspond to the
famous dominant and scaling attractors respectively which are also
found in quintessence model\cite{14, 35} and tachyon model\cite{15,
151}. Similar to \textit{cp$6$}, \textit{cp$7$} is also a dark
energy dominated attractors ($\Omega_\phi=1$). However,
\textit{cp$7$} is different from \textit{cp$6$} on two points. The
first difference is that \textit{cp$7$} is not a de-Sitter-like
dominant attractor since the state equation $w_\phi$ of
\textit{cp$7$} could be any value between $-1$ and $-1/3$(see
Fig.\ref{fig3}, note that $0\leq\gamma_\phi=w_\phi+1\leq2/3$)
depending on the different value of $\lambda_\ast$. So the value of
state equation of dark energy $w_\phi$ could be matched to the value
from the observational data. The second difference is that the sound
speed $c_s^2$ is monotonically increasing with $\gamma_\phi$ instead
of being zero(see Fig.\ref{fig4}). \textit{cp$8$} is the scaling
solution that power-law kinetic quintessence will track the
evolution of background matter in the early time and behaves as the
ordinary matter with $\Omega_{\phi}=3/\lambda_\ast^2$ and
$\gamma_\phi=1$. We proved in section 4 that this scaling is only
possible for ordinary matter($\gamma_b=1$), which is different from
the result obtained in quintessence model\cite{14, 35}.

\par Though the existence and stable conditions are different for \textit{cp$6$},
\textit{cp$7$} and \textit{cp$8$}, it is very interesting to
consider the possibility that our universe may evolve continuously
from one stable critical point(\textit{cp$8$}, scaling solution) to
another stable one(\textit{cp$6$} or \textit{cp$7$}, dark energy
dominant solution). We borrow the idea in Ref\cite{16} where the
author proposed a scenario of universe which could evolve from a
scaling attractor to another dark energy dominant attractor by
introducing a field whose value changed by a certain amount in a
short time. Since the change of the value of scalar field $\phi$
means the change of the value of $\lambda_\ast$, then
\textit{cp$6$}, \textit{cp$7$} and \textit{cp$8$} could be stable
before and after the change of scalar field $\phi$. Actually, we can
also obtain these two asymptotical evolutions if the potential
$V(\phi)$ can be approximated to two different potentials when
scalar field $\phi$ evolves into different ranges: one admits the
scaling solution and another admits the dark energy dominant
solution\cite{351, 352, 353}. For these potentials, the exit of the
cosmological evolution from one attractor solution to another
attractor is quite natural, but the explanation of why we have these
special potentials may require fine tuning.

\par In the end of this paper, we would like to discuss \textit{cp$7$} and \textit{cp$8$} from the observational point of
view. The sound speed $c_s^2$ in the case of \textit{cp$7$} and
\textit{cp$8$} is very special comparing to other scalar field
models, which makes it possible to distinguish the power-law kinetic
quintessence from other scalar field models using the observational
data. The sound speed $c_s^2$ of \textit{cp$7$} is monotonically
increasing with $\gamma_\phi$ instead of being zero. We can see from
Fig.\ref{fig4} that $0\leq c_s^2\leq1/9$ when
$0\leq\gamma_\phi\leq2/3$, and the closer $w_\phi$ is to $-1$, the
closer $\gamma_\phi$ is to $0$. We also plot the relationship
between $c_s^2$ and $\gamma_\phi$ of canonical quintessence and
tachyon in Fig.\ref{fig4} for comparing with each other. We can find
that the value of $c_s^2$ of power-law kinetic quintessence is
dramatically less than the sound speed of canonical quintessence and
tachyon. Since the nature of dark energy can be probed not only
through $w_\phi$ but also through its microphysics, characterized by
the sound speed of perturbations to the dark energy density and
pressure. As the sound speed $c_s^2$ drops below the speed of light,
dark energy inhomogeneities increase, affecting both cosmic
microwave background and matter power spectra\cite{9, 37}. It is
shown that observational data may distinguish the dark energy models
with the sound speed $c_s^2\ll1$ from the models with
$c_s^2\rightarrow1$\cite{10, 381, 39, 40, 41, 42}, so in principle
it could be distinguished from canonical quintessence and
tachyon(see Fig.\ref{fig4}). For the case of \textit{cp$8$}, power
law kinetic quintessence behaves as ordinary matter with
$\gamma_\phi=1$. For the ordinary matter(or dark matter), we know
that $c_s^2=0$. However, it is very interesting to notice here that
the sound speed $c_s^2$ equals $1/5$ from Eq.(\ref{eq81}). It means
that power-law kinetic quintessence track the evolution of ordinary
matter with the same state equation $w_m$ but a different sound
speed $c_s^2$. This is very important since the behavior of
perturbation in scalar field dark energy and its consequent effect
on the cold dark matter power spectrum is governed by the state
equation $w_\phi$ and the effective speed of sound $c_s^2$ of dark
energy. For the scaling solution \textit{cp$8$}, the dark energy
density was non-negligible($\Omega_{\phi}=3/\lambda_\ast^2\neq0$) at
early times(Early Dark Energy models\cite{43}). It is shown that as
$\gamma_\phi$ gets further from 0, the influence of the sound speed
increases; for models with $\gamma_\phi\approx 1$ at high redshift
there is also the possibility of non negligible amounts of early
dark energy density. So even just a couple percent of the total
energy density in early dark energy can dramatically improve the
prospects for detecting dark energy clustering\cite{37}. The impact
of early dark energy fluctuations in both linear and nonlinear
regimes of structure formation had been explored in Ref\cite{44}. In
these models the energy density of dark energy is non-negligible at
high redshift and the fluctuations in the dark energy component can
have the same order of magnitude of dark matter fluctuations.
However, how the impact will be changed if the $c_s^2$ equals $1/5$
for power-law kinetic quintessence is not investigated yet. Since
power-law kinetic quintessence are special both in the early
universe where it is an early dark energy tracking the ordinary
matter with $c_s^2=1/5$ and in the late universe where it drives the
accelerating expansion with $c_s^2\rightarrow0$, it may be a good
suggestion to simultaneously investigate the perturbations of dark
energy at both early and late time universe, and to explore whether
there are any degeneracies of the impacts between early dark energy
and late dark energy on cold dark matter power spectrum and cosmic
microwave background.

\section{Acknowledgement}
\par This work is partly supported by National Nature Science Foundation of China under Grant No. 11333001, Shanghai Research Grant No. 13JC1404400 and Shanghai Normal University
Research Program under Grant No. SK201309.  \\

 {\Large \bf Appendix}
\par {\small In section 4, we pointed out that if the eigenvalues of
Jacobi matrix has one or more eigenvalues with zero real parts while
the rest of the eigenvalues are negative, then linearization fails
to determine the stability properties of this critical point. Among
the six critical points(\textit{cp$1$}, \textit{cp$3$},
\textit{cp$4$}, \textit{cp$6$} ,\textit{cp$7$}, \textit{cp$8$})
investigated in this paper, \textit{cp6} is just such point. So in
this Appendix we will show you how we get the stable condition of
\textit{cp6} using the center manifold theorem\cite{34}. The point
\textit{cp6} is: $(\Omega_{\phi}, \gamma_{\phi}, \lambda)=(1, 0,
0)$, and its three eigenvalues are $-3\gamma_b$, $-3$ and $0$.
Firstly, we need to transfer \textit{cp6} to $cp6'$
$(\Omega_1=\Omega_{\phi}-1, \gamma_{\phi}, \lambda)=(0, 0, 0)$ for
convenience. In this case, Eqs.(\ref{eq5}, \ref{eq6}, \ref{eq10})
can be rewritten as:

\begin{equation}\label{A1}\frac{d \Omega_1}{dN}=-3(\gamma_b-\gamma_{\phi})\Omega_1(\Omega_1+1)\end{equation}
\begin{equation}\label{A2}\frac{d \gamma_{\phi}}{dN}=\frac{(\lambda\sqrt{3(4-3\gamma_{\phi})(\Omega_1+1)}+3\gamma_{\phi})(\gamma_{\phi}-2)(3\gamma_{\phi}-4)}{3\gamma_{\phi}-8}\end{equation}
\begin{equation}\label{A3}\frac{d\lambda}{dN}=\lambda^2\sqrt{3(4-3\gamma_{\phi})(\Omega_1+1)}~f(\lambda)\end{equation}

Noted that now $\Omega_1$, $\gamma_{\phi}$, $\lambda$ in
Eqs.(\ref{A1}, \ref{A2}, \ref{A3}) are very small variables around
$cp6'$ ($\Omega_1=0, \gamma_{\phi}=0, \lambda=0$). Function
$f(\lambda)$ in Eq.(\ref{A1}) could be taken the taylor series in
$\lambda$:
$f(\lambda)=f(0)+f^{1}(0)\lambda+\frac{f^{2}(0)}{2!}\lambda^2+...$,
where  $f^{n}(0)$ is the value of $\frac{d^n f(\lambda)}{d
\lambda^n}$ when $\lambda=0$.

\par We get the Jacobi matrix ${\cal A}$ of nonlinear autonomous dynamical
system Eqs.(\ref{A1}, \ref{A2}, \ref{A3}) from Eq.(\ref{eqA}):

\begin{equation}\label{A4}{\cal A}= \left [ \begin{array}{lll} -3\gamma_b & \ \ 0 & \ \ \ \ 0 \\ \ \ 0 \ & -3 &\ \ -2\sqrt{3}
\\ \ \ 0 &\  \ 0 & \ \ \ \ 0\end{array} \right ]\end{equation}

\par The eigenvalues of ${\cal A}$ and the corresponding eigenvectors are:

\begin{equation}\label{A5} \{-3\gamma_b,\ \ [1, 0, 0]\};\ \ \ \ \ \{-3,\ \  [0, 1, 0]\}; \ \ \ \ \ \{0,\ \ [0,-\frac{2}{\sqrt{3}}, 1]\}\end{equation}

Let ${\cal M}$ be a matrix whose columns are the eigenvectors of
${\cal A}$, then we can write down ${\cal M}$ and its inverse matrix
${\cal T}$:

\begin{equation}\label{A6}{\cal M}= \left [ \begin{array}{lll}\ 1 & \ 0 & \ \ 0 \\ \ 0 & \ 1 & -\frac{2}{\sqrt{3}} \\  \ 0 & \ 0 & \ \ 1\end{array} \right ],
 \ \ \ \  \ \ {\cal T}={\cal M}^{-1}= \left [ \begin{array}{lll}\ 1 & \ 0 & \ \ 0 \\ \ 0 & \ 1 & \frac{2}{\sqrt{3}} \\  \ 0 & \ 0 & \ \
1\end{array} \right ]\end{equation}

Using the similarity transformation ${\cal T}$ we can transform
${\cal A} $ into a block diagonal matrix, that is,

\begin{equation}\label{A7} {\cal T}{\cal A}{\cal T}^{-1}=\left [ \begin{array}{lll} -3\gamma_b & \ 0 & \ \ 0 \\ \ \ 0 & -3 &\ \ 0 \\ \ \ 0 & \ 0 & \ \ 0
\end{array} \right ]=\left [\begin{array}{ll}{\cal A}_1 & \ 0 \\ 0 &  {\cal A}_2 \end{array} \right ]\end{equation}
where all eigenvalues of ${\cal A}_1$ have negative real parts while
eigenvalue of ${\cal A}_2$ has zero real part. We then change the
variables ($\Omega_1, \gamma_{\phi}, \lambda$) in Eqs.(\ref{A1},
\ref{A2}, \ref{A3}) to another set ($\Omega_2, \gamma_1, \lambda_1$)
as follows:

\begin{equation}\label{A8} \left [\begin{array}{l}\Omega_2 \\ \gamma_1 \\ \lambda_1\end{array} \right ] ={\cal T} \left [\begin{array}{l}\Omega_1 \\
 \gamma_\phi \\ \ \lambda\end{array}\right ]=\left [\begin{array}{l}\ \ \ \Omega_1 \\ \gamma_\phi+\frac{2}{\sqrt{3}}\lambda \\ \ \ \ \ \lambda\end{array}\right ]\end{equation}

 Then we can rewrite the dynamical system Eqs.(\ref{A1}, \ref{A2}, \ref{A3}) in the form of the new variables:

\begin{equation}\label{A9}\frac{d\Omega_2}{dN}=\frac{d\Omega_1}{dN}=-3(\gamma_b-\gamma_1+\frac{2}{\sqrt{3}}\lambda_1)(\Omega_2+1)\Omega_2=G_1(\Omega_2, \gamma_1, \lambda_1)\end{equation}
$$\frac{d\gamma_1}{dN}=\frac{d\gamma_\phi}{dN}+\frac{2}{\sqrt{3}}\frac{d\lambda}{dN}=2\lambda_1^2\sqrt{(4-3\gamma_1+2\sqrt{3}\lambda_1)(\Omega_2+1)}f(\lambda_1)$$
\begin{equation}\label{A10}+\frac{(\lambda_1\sqrt{3(4-3\gamma_1+2\sqrt{3}\lambda_1)(\Omega_2+1)}+3\gamma_1-2\sqrt{3}\lambda_1)(\gamma_1-\frac{2}{\sqrt{3}}\lambda_1-2)(3\gamma_1-2\sqrt{3}\lambda_1-4)}{3\gamma_1-2\sqrt{3}\lambda_1-8}=G_2(\Omega_2, \gamma_1, \lambda_1)\end{equation}
\begin{equation}\label{A11}\frac{d\lambda_1}{dN}=\frac{d\lambda}{dN}=\lambda_1^2\sqrt{3(4-3\gamma_1+2\sqrt{3}\lambda_1)(\Omega_2+1)}~f(\lambda_1)=G_3(\Omega_2, \gamma_1, \lambda_1)\end{equation}

\par According to the center manifold theorem, the stable properties of \textit{cp6} is determined by reduced system Eq.(\ref{A11}).
We set the center manifold for $\Omega_2$ and $\gamma_1$:

\begin{equation}\label{A12}\Omega_2=h_1(\lambda_1); \ \ \ \ \ \gamma_1=h_2(\lambda_1)\end{equation}
where $h_1, h_2$ are the function of $\lambda_1$ which satisfy
following conditions:

\begin{equation}\label{A13}h_1(0)=0; \ \ \frac{\partial h_1}{\partial \lambda_1}=0;\ \ \ \ h_2(0)=0; \ \frac{\partial h_2}{\partial \lambda_1}=0\end{equation}

The center manifold equations are as follows:

\begin{equation}\label{A14}G_1(h_1(\lambda_1), h_2(\lambda_1), \lambda_1)=\frac{\partial h_1}{\partial \lambda_1} \cdot G_3(h_1(\lambda_1), h_2(\lambda_1), \lambda_1)\end{equation}
\begin{equation}\label{A15}G_2(h_1(\lambda_1), h_2(\lambda_1), \lambda_1)=\frac{\partial h_2}{\partial \lambda_1} \cdot G_3(h_1(\lambda_1), h_2(\lambda_1), \lambda_1)\end{equation}

We set $h_1=a_2\lambda_1^2+a_3\lambda_1^3+\cdots,
h_2=b_2\lambda_1^2+b_3\lambda_1^3+\cdots$ and substitute these
series into the center manifold equations Eqs.(\ref{A14}, \ref{A15})
to find the unknown coefficients $a_2, b_2, a_3, b_3, \cdots$ by
matching the coefficients of like powers in $\lambda_1$ in the left
and right sides of equations Eqs.(\ref{A14}, \ref{A15}). However, we
do not know in advance how many terms of the series of $h_1, h_2$ we
need. We start with the simple approximation:
$h_1=a_2\lambda_1^2+a_3\lambda_1^3,
h_2=b_2\lambda_1^2+b_3\lambda_1^3$, and get that $a_2=a_3=0$,
$b_2=-\frac{1}{2}+\frac{4}{3}f(0)$, $b_3=
\frac{4}{3}f^{1}(0)+\frac{5}{9}\sqrt{3}f(0)-\frac{\sqrt{3}}{16}-\frac{16}{9}\sqrt{3}f(0)^2$.
We therefore set
$h_1=a_4\lambda_1^4+a_5\lambda_1^5+a_6\lambda_1^6+a_7\lambda_1^7$,
and found $a_4=a_5=a_6=a_7=0$. So we finally set $h_1=0$,
$h_2=[-\frac{1}{2}+\frac{4}{3}f(0)]\lambda_1^2+[\frac{4}{3}f^{1}(0)+\frac{5}{9}\sqrt{3}f(0)-\frac{\sqrt{3}}{16}-\frac{16}{9}\sqrt{3}f(0)^2]\lambda_1^3$.

So we substitute $h_1, h_2$ into the reduced system Eq.(\ref{A11})
and rewrite it as follows:

\begin{equation}\label{A16}\frac{d\lambda_1}{dN}=\lambda_1^2\sqrt{3}\sqrt{4+2\sqrt{3}\lambda_1-[4f(0)-\frac{3}{2}]\lambda_1^2-[\frac{5}{3}\sqrt{3}f(0)-\frac{16}{3}\sqrt{3}f(0)^2+4f^1(0)-\frac{3\sqrt{3}}{16}]\lambda_1^3}~f(\lambda_1)\end{equation}

 where $f(\lambda_1)$ can be expanded as
 $f(0)+f^{1}(0)\lambda+\cdots$. Since $\lambda_1$ is a very small variable around $\lambda_1=0$, so Eq.(\ref{A16}) can be simplified in the neighborhood
 of  $\lambda_1$ as follows:

\begin{equation}\label{A17}\frac{d\lambda_1}{dN}=2\sqrt{3}f(0)\lambda_1^2\end{equation}
where $f(0)$ is the value of function $f(\lambda)$ at $\lambda=0$.
The stability of $\textit{cp6}$ will be finally determined by above
simplest reduced system Eq.(\ref{A17}). It is clear that the stable
condistion for dynamical system Eq.(\ref{A17}) is

\begin{equation}\label{A18} f(0)<0 \end{equation}

\par So in this Appendix we proved that $\textit{cp6}$ is a stable de-Sitter-like dominant
attractor when $f(0)<0$, just as stated in section 4.
\\

\end{document}